\documentstyle[12pt]{article}
\def\beq{\begin{equation}}
\def\eeq{\end{equation}}

\def\bea{\begin{eqnarray}}
\def\eea{\end{eqnarray}}
\def\ci{\cite}
\def\la{\label}

\begin{document}

\begin{flushright}
hep-ph/9411443 \\
HD-THEP-94-45  \\
IFUNAM-FT-94-63 \\

\end{flushright}
\vspace{15mm}

\begin{center}
{\bf INFLATION FROM SUPERSTRINGS
} \\
\end{center}

\vspace*{0.7cm}

\begin{center}
{\bf A. De la Macorra$^{a}$ and S. Lola$^{b}$}
\end{center}

\vspace*{0.1cm}
\begin{center}
\begin{tabular}{c}
$^{a}${\small Instituto de Fisica, UNAM}\\
{\small Apdo. Postal 20-364,
01000  Mexico
D.F., Mexico}\\
$^{b}${\small Institut f\"{u}r
Theoretische Physik, Univerisit\"at
Heidelberg,}\\
{\small Philosophenweg 16, 69120 Heidelberg, Germany }\\
\end{tabular}
\end{center}

\vspace{1 cm}

\begin{center}
{\bf ABSTRACT}
\end{center}

{\small We investigate the possibility of obtaining
inflationary solutions of the slow roll type from a low
energy Lagrangian coming from
superstrings. The advantage of such an approach is that
in these theories the scalar potential has only one free
parameter (the Planck scale) and therefore
no unnatural fine tuning may be accommodated.
We find that in any viable scheme the dilaton and the moduli
fields have to be stabilized and that before
this happens, no other field may be used
as the inflaton.
Then inflation may occur due to
chiral matter fields. Demanding that the potential
terms associated with the chiral fields
do not spoil the dilaton and moduli minimization
leads to severe constraints on the magnitude of the density fluctuations.
}

\noindent
\rule[.1in]{14.5cm}{.002in}

\thispagestyle{empty}

\setcounter{page}{0}
\vfill\eject

\section{Introduction}

The standard hot big-bang theory,
although in general successful,
has several shortcomings \cite{standard}.
Among them are the flatness and horizon
problems, the overabundance of topological
defects if a GUT symmetry has existed, as
well as the origin of the density fluctuations
that have lead to galaxy formation \cite{faults}.
It has been found that
these problems may be addressed
if the
universe in its very early stages has
been in an unstable vacuum- like state
\cite{inflation}.
In this case the scale factor $R$
grows exponentially
till the energy stored in the
vacuum transforms into thermal energy.
Subsequently
the universe is described by the
standard theory. At the end of the period of the
exponential growth of the scale factor (inflationary era),
density fluctuations
given by \cite{fluc}
\begin{equation}
\frac{\delta \rho}{\rho}
=
C \, \left. \frac{H(\Phi) \delta \Phi}{ \dot{\Phi}}
\right|_{k \sim H}
\label{eq0}
\end{equation}
are to be expected
for a scalar field $\Phi$
whose potential energy dominates\footnote{In the
following we will set the reduced Planck
mass $m_p$
 to one (i.e. $m^2_p=M^2_{Planck}/8\pi=1$).} and $C\sim O(1)$ \ci{C.hotcold}.

Since the COBE
measurements of the cosmic
microwave background radiation
which gave evidence of primordial fluctuations
\cite{COBE}, the
interest in inflationary theories has been
revived.
And although there exists a discrepancy
between the COBE observations and
the existing cold or hot dark matter
models for structure formation
\cite{p3}, it is possible
to reconcile theory with observation,
by considering either a combination of
hot and cold dark matter,
or by including
the effect of additional
sources of fluctuations.
For example, we have shown
that, unlike what was previously thought,
under certain conditions
domain walls may enhance the standard
cold dark matter spectrum without
inducing unacceptable cosmic microwave
background distortions.
This occurs provided that either
one of the minima of the potential of the scalar
field $\Phi$ is favoured \cite{wa2}, or
the domain walls are unstable and annihilate
after having induced fluctuations to the
cold dark matter background \cite{wa1}.

Among the inflationary solutions
one of particular interest is chaotic
inflation \cite{chao}. According
to this theory the initial
distribution of a scalar field
is random at the many
causally disconnected
regions which correspond to the
horizon today.
The domain where the initial conditions of the scalar
field are consistent with inflation will then
expand rapidly and cover the whole
of the visible universe, while the rest of the
domains will remain frozen. The advantage of this approach
is that no fine-tuning associated with the initial conditions
is necessary.
However,
although this is an attracting idea,
the existing schemes
share a common problem:
in order to obtain
sufficient growth of the scale factor
as well as the correct magnitude
of density fluctuations,
an unnatural fine tuning of the
models including the introduction of
a very small coupling constant in the
theory is required.

This unnatural fine tuning may be avoided,
by going in models where
a transition from a higher dimensional to a four-dimensional
universe occurs \cite{wett}. Then, the models
are governed by
only one scale, the Planck
mass $M_{Planck}$,
and the parameters that lead to
successful inflation
need not be tuned to tiny values.
The best candidate of a higher dimensional
unification is superstring theory.
Since all the parameters of the theory
below the Planck scale
are dynamically determined and
no fine tuning may be accommodated,
inflation in low energy
models coming from superstrings
either works in a natural
way or does not work at all.
Deriving an
inflationary
scalar potential in string models has been
difficult \cite{stein},
mainly due to the dilaton field, and
until now
no viable example has been obtained.
There are several proposals on inflation
from superstrings \ci{stew} of the slow roll type,  however,  either
they do not  deal with the dynamics of the
dilaton field or they violate some of
the "current superstring lore"\footnote{For an alternative solution
to inflation see \ci{Venez}. In this case inflation is due  to the
kinetic energy  of the inflaton and not to the scalar potential.}.

In the following analysis we are going to
look for possible inflationary solutions
in the framework of low
energy superstring theory.
In section 2
we give a short outline of a low energy
Lagrangian coming from the heterotic superstring
compactified on
orbifolds.
In section 3
we start our investigation
by examining whether the dilaton
alone may induce inflation.
We find that this is
possible  neither at tree level,
nor when loop corrections are included.
The same is true for the moduli fields.
Moreover, unless the dilaton and the moduli
are stabilised, no other field may be used as
the inflaton.
Section 4 deals with chiral fields.
In 4.1 we first
look at the conditions for
inflation for chiral fields.
In 4.2
we give the conditions that the potential
of the chiral fields has to obey so
as not to spoil the
dilaton and the moduli minimization.
This implies a natural
constraint on the magnitude of the
post-inflationary fluctuations,
a novel feature so far.
In 4.3
we consider the region with
small values for the
chiral fields and we find that
in this limit no viable
solution exists. In 4.4 we examine the
generic behaviour of
the potential for
large values of the chiral fields
and we show that under certain conditions
it is possible to
obtain inflation.
In  section 5 we describe a viable
solution. In particular, in 5.1
we derive the scalar
potential and its derivatives for
the viable scheme, while in
5.2 we calculate
the number of e-folds of
inflation as well as the
magnitude of the density fluctuations.
Finally, in section 6 we give a summary
of our results.

\section{4D-Superstring theory}

We work with a low energy Lagrangian of
the heterotic superstring \cite{effective} which
has been
derived by
orbifold compactification.
The effective $ D=4 $
superstring  model
is given by an $N = 1$
supergravity  theory \cite{r15}
with at least  four gauge singlet fields
$S$ and $T_i,  \,i=1,2,3$ as well as an
unspecified number of gauge
chiral matter superfields. The v.e.v. of the dilaton field $S$ gives
the gauge
coupling constant $g^{-2} = {\rm Re} S$ at the string scale
 while the real part of the moduli fields $Re\, T_i= R^2_i$ the
radius of the
compactified dimension.
The tree level scalar potential
is in general  expressed by \ci{r15}
\bea
V_{0} &=& \frac{1}{4} e^{K} f
\\
f &=& (G_{a} (K^{-1})^{a}_{b} G^{b} -3|W|^{2})
\label{eq1}\eea
where
\begin{eqnarray}
G_{a} & \equiv &
K_{a}W + W_{a} \nonumber \\
K & = & - \log(S_{r}) -
\log \left[ (T_{r} -\Phi\bar{\Phi})^3 - B \bar{B} -
T_{r} (C\bar{C}) \right ]
\label{e2}\\
W & = &  W_0(S,T) + P(T,\Phi,B,C). \nonumber
\end{eqnarray}
$G$ is the K\"{a}hler
potential,
$T_{r} =   T + \bar{T}, S_{r}  =  S + \bar{S} $ and the indices
$a,b$ run
over all chiral fields, i.e. the dilaton $S$, the moduli $T$ and
chiral fields
$\Phi$, $B$ and $C$. The $\Phi$ fields correspond to untwisted chiral
fields
while $B$ and $C$ are twisted fields which appear naturally in
orbifold
compactification. We have consider for simplicity an overall moduli
$T$ (we
will take   different $T_i$ fields when necessary). All the fields
are
normalized with respect to the reduced Planck mass
 $m_{p} =  M_{p}/\sqrt{8 \pi}$.
The term $W_0(S,T)$ arises due to non-perturbative
effects, like gaugino condensation\cite{r9} and is responsible
for
breaking supersymmetry (SUSY), while $P$ is
the chiral matter
superpotential.
In
particular $P$ contains the trilinear (Yukawa) interactions of the
chiral
fields. As usual, the indices $a,b$
of the functions $K$ and $W$ denote derivatives with respect to
chiral fields.

The form of $K$ is derived by a perturbative
expansion and is valid if the arguments inside
the logarithms are positive. This indicates that
\begin{eqnarray}
\, 0 & \leq &  S_{r} \nonumber \\
\Phi \bar{\Phi} & \leq & T_{r}^{n}
\label{eq3}
\end{eqnarray}
where we wrote
$\Phi$, $B$ and $C$ in a unified
way by using
the index $n$ that runs
from $1$ to $3$, thus
taking into account the
different modular weights  of the
three distinct chiral
fields with respect to $T_{r}$.
These inequalities indicate that
there is a limit to the range that we can
explore in the framework of the current models.
The above expression for $K$ is the one at tree-level.
When one includes loop effects a mixed term between the dilaton and
the moduli
fields term may arise
so that  $K_{S}^{T} \neq 0$. In this
paper  we will not consider the mixed
$S, T$ term in $K$ because  one
expects it
to be much smaller than the tree level
contributions
and will therefore not
affect the
analysis done here.

In the subsequent sections we
investigate whether
we can get inflation from
the low energy superstring potential
and we initially concentrate on
the dilaton field.

We work with the effective
low energy superstring Lagrangian in the Einstein
and not in the Brans-Dicke frame \ci{Bellido} which would seem a
priori more natural in the context of string inflation. However,
most of the work in determining the  non-perturbative contributions
of S and T to the potential,  e.g. SUSY breaking terms (like gaugino
condensation) or the study of the duality symmetries  for $T$ and
$S$,  has been carried out in the  Einstein frame.

\section{Dilaton Field $S$}

The dilaton field $S$ is present in all 4-D string models   and its
interaction
with other chiral fields is generic.  Since a
successful
inflationary potential  must inflate due to all
dynamical fields,
unless they
are at their minimum, it is necessary to determine whether the scalar
potential
inflates due to the dilaton field.

For   the K\"{a}hler potential $K$ given in eq.(\ref{e2}) we have
\begin{equation}
K_S= \frac{\partial K}{\partial S}
= - \frac{1}{S_r}, \,\,\,
K_{S}^{S} = \frac{\partial^{2} K}
{\partial S \; \partial \bar{S}} =
\frac{1}{S_{r}^{2}}.
\label{eq4}
\end{equation}
and
\beq
K_{S}(K^{-1})_{S}^{S}K^{S} = 1.
\eeq
We consider first the  perturbative  superpotential $W$  which has no
$S$
dependent  terms.  In this case
 the interaction of
$S$ with the other scalar fields is through a
potential of the form
\beq
V = \frac{1}{S_{r}}f
\eeq
where $f$ (cf. eq.(\ref{eq1})) is now independent of $S$.
Then
\beq
V_{S}  \equiv
\frac{\partial V}{\partial S} = -\frac{V}{S_{r}}.
\eeq
Using the Lagrangian density for $S$
\begin{equation}
L = R^{3} \left ( K_{S}^{S} \, \partial_{i} S
\, \partial^{i} \bar{S} - V \right )
\label{eq4a}
\end{equation}
where $R$ is the scale factor,  the equation of motion for $S$
is (when ignoring
the second derivative terms with
respect to time)
\begin{equation}
3 H K_{S}^{S} \dot{S}+ \dot{K_{S}^{S}} \dot{S}
= -\frac{\partial V}
{\partial S}
\label{eq5}
\end{equation}
and a similar equation is obtained for $\bar{S}$.
By considering $V$ to be constant
the Einstein's equations give the
relation
\begin{equation}
H^{2} \equiv \frac{\dot{R}^{2}}{R^{2}} = \frac{V}{3}.
\label{eq5b}
\end{equation}
Substituting eq.(\ref{eq5b}) back to
eq.(\ref{eq5}) and using eq.(\ref{eq4})
gives the number of e-folds of inflation
\beq
N =  -\int \frac{K_{S}^{S}V}{V_{S}}dS =
\ln
\left (
\frac{S_{r_{e}}}{S_{r_{i}}}
\right )
\label{x}
\eeq
where the subindices ``i''
and ``e''
stand for the beginning and the end
of inflation respectively{\footnote{Here we have dropped the term
$\dot{K}_{S}^{S} \dot{S}$ which is smaller
than $3 H K_{S}^{S}$.}.
For enough inflation, $N$ has to be
$ \geq 65$, thus
$S_{r_{i}}/S_{r_{e}} \approx e^{65} \gg 1$.
Since satisfying this condition in the framework
that we have discussed so far is quite unnatural,
one cannot have inflation
for the dilaton field without an $S$-dependent superpotential.

A non-perturbative $S$-dependent superpotential may be generated
by a strong gauge coupling constant that leads to the formation of a
gaugino condensate \cite{r9},
\ci{gau}, \ci{axel}.
To describe the interaction of the
gaugino condensate with other
fields
one introduces a scalar field $U$,
which, after using the equations of motion,
is expressed in terms of the  gaugino
bilinear ($\lambda \bar{\lambda}$).  Its interaction is determined by
symmetries
and anomaly cancellation arguments. After the $U$ field has been
integrated out,
an effective
superpotential for the dilaton field $S$  is generated
\begin{equation}
W_0 = h e^{-3 S /2 b_0}
\sim \Lambda_{c}^{3}
\label{eq7}
\end{equation}
where $b_{0}$ is the one-loop
coefficient
of the
beta function
of the hidden sector gauge
group,
$h$ a coefficient which is
independent of $S$ but may depend on the moduli fields $T$
and
$\Lambda_{c}$ is the condensation scale.
The tree level scalar
potential, for $h$ independent of $T$,  is
given by
\beq
V_0 =  \frac{1}{4}
e^{K}
\left(
|W-S_r W_{S}|^{2}
+ (K_{i}(K^{-1})^{i}_{j}K^{j}-3)
|W|^{2}
\right)
\eeq
with $i,j$ standing
for derivatives with respect to
chiral fields $T_i, \Phi_{i}$.
Since for a large hierarchy solution (i.e.
the masses of the scalar
fields are
much smaller then the Planck mass) one has
$b_{0} =
\frac{3N-n_{f}/2}{16 \pi^{2}} \ll 1$,
the
$|W_{S}|^2$ term dominates the potential $V_{0}$,
\begin{equation}
V_{0}
\approx   \frac{1}{4} e^{K}(K^{-1})_{S}^{S}
|W_{S}|^{2} =  \frac{1}{4} h^{2} e^{K_0}e^{-3 S_{r}/2 b_0}S_{r}
\label{eq8}
\end{equation}
where we have written $K = -\ln{S_{r}} + K_{0}$ with $K_{0}$ being
S-independent.

In this case the number of e-fold of inflation are
\beq
|N| = |-\int \frac{K_{S}^{S}V}{V_{S}}dS|
\approx \frac{2 b_0}{3
S_{r}} \ll 1
\label{x2}
\eeq
where $V_{S}$ is the first derivative
of the potential with respect to
$S$.
As we see the number of e-folds
is too small to have any
effect on the evolution of the
universe.
This indicates that even  with the non-perturbative superpotential
$W_0$ there
is no inflation.

However, the potential described in eq.(\ref{eq8})
does not have a stable solution
in the
dilaton direction. It is unbounded from
below for $S \rightarrow 0$
and it goes
to zero for $S \rightarrow \infty$.
Up to now, there are two possibilities to stabilize the potential
with a large
hierarchy\footnote{It has also been suggested to consider an
S-dual
invariant potential to stabilize the dilaton
\cite{Sdual}. We leave the study of
the S-dual
invariant inflationary potential for a feature work.} :\\
(i) to consider a single gaugino condensate and include
loop corrections of the effective 4-Fermi gaugino
interaction \cite{axel}\\
(ii) to introduce two gaugino condensates
with slightly different one-loop
beta function coefficients \ci{gau}.\\
An important
difference between the two
approaches is  the v.e.v. of the moduli.  For the two gaugino
condensation case
$<T > \simeq 1$ while for the single gaugino
condensation one finds
that $<T>$
can be much larger,  $<T>=O(10-25)$, allowing for an unification
scale of the
order of $10^{16}\,GeV$ \ci{uni} as required by the minimal
supersymmetric
standard model \ci{mssm}.

We are now going to examine whether   the dilaton potential
may induce inflation.
Let us  consider, for simplicity,  three different regions
for the dilaton:

\underline{(i) around the minimum}

In this region
\beq
V_{S}|_{S_{r_{0}}} = V|_{S_{r_{0}}} = 0.
\eeq
Here we are assuming that the minimum has vanishing cosmological
constant\footnote{Without this assumption $V$ is negative at the
minimum and
the universe would  clearly not inflate.} (we  refer to the
cancellation of the
cosmological constant at the end of section 4.4).
By expanding $V$ and $V_{S}$ around the minimum  the leading term is
given in
terms of the
 second derivative of $V$ and
\beq
\frac{K_{S}^{S}V}{V_{S}} \approx
\frac{K_{S}^{S}V_{SS}(S_r-S_{r0})^2/2}{V_{SS}(S_r-S_{r0})}
= \frac{1}{2}
\frac{(S_{r}-S_{r_{0}})}{S_{r}^{2}}.
\eeq
The number of e-folds is
\beq
N = -\int \frac{K_{S}^{S}V}{V_{S}}dS =
\frac{1}{2} \ln \left
(\frac{S_{r_{e}}}{S_{r_{i}}}
\right) + \frac{S_{r_{0}}}{2}
\left (
\frac{1}{S_{r_{e}}} -\frac{1}{S_{r_{i}}}
\right)
\eeq
which indicates that for small
perturbations around $S_{r_{0}}$ the potential does not
inflate.
This is a generic result, saying that
no potential with $V_{S}|_{S_{r_{0}}}$
= $V|_{S_{r_{0}}} =0$ and $K_{S}^{S} = 1/S_{r}^{2}$
may inflate enough around the minimum.

\underline{(ii)  $S_{r} < S_{r_{0}}$}

For $S_{r}$ away from the minimum
$V$ behaves as $V \approx \frac{1}{S_{r}}e^{-S_{r} a} + ...$,
$a$ being a constant,
therefore if $S_{r} < S_{r_{0}}$
\beq
V \sim \frac{1}{S_{r}}
\eeq
which as we have seen does not
lead to inflation (cf. eq(\ref{x})).

\underline{(iii) $S_{r} > S_{r_{0}}$}

In this case
\beq
V \sim e^{-S_{r} a}
\eeq
and again no inflation occurs (cf. eq.(\ref{x2})).

Thus, we have shown that the dilaton field does  not induce
inflation. Furthermore,  the
$S$ and $U$  fields
are  stabilized at the same time, so
we think that
it is not possible
to use $U$ as the inflaton.

The moduli field, $T$, has a similar behaviour
 to the dilaton field, that is
they have an $1/T$ or an exponential
potential,  and
 therefore it seems difficult that the universe could inflate due to
this
field. However, since the Yukawa couplings and many other terms in
the
superpotential are in general moduli dependent,
a more complicated
dynamics
arises for the moduli than for the dilaton.  It could happen that
some
cancellations  in the scalar potential take place allowing for
inflation. However we
consider this possibility
not very plausible and at least very much
model
dependent, so we will assume that inflation is not
occurring due to the moduli
fields either.

{}From the above discussion  we would
like to emphasise the following points:

$\bullet$
Inflation should occur for
{\em all} dynamical fields that are not at their
minimum, so unless  $S$ and $T$ are  stabilized
{\em no other scalar field may
be used as the inflaton}.

$\bullet$
$S$, $U$ and $T$ become fixed at the same time \ci{axel}
therefore if inflation occurs, it should be
due to {\em chiral matter fields} and {\em below} the supersymmetry
breaking
scale, where $S, T$ and $U$ are stabilized.

\section{Chiral fields}

We now pass to the chiral matter fields,
which have to be considered in combination
with the moduli fields since the K\"{a}hler potential introduces
mixing
terms
between  them both   (cf. eq.(\ref{e2})).

\subsection{Conditions for inflation for chiral fields}

We first look at the conditions for inflation
for chiral fields with non-canonical kinetic terms
(that is for $K_{i}^{j} \neq \delta_{i}^{j}$).

The Lagrangian density is given by
\beq
{\cal L} = (-g)^{1/2}
\left (
\frac{1}{2}g^{\mu\nu}K_{i}^{j}
\partial_{\mu}
\Phi_{j}
\partial_{\nu}
\Phi^{i}-V(\Phi_{i}) \right),
\label{lag}
\eeq
where $(-g)^{1/2} = R^{3}$
and
$g_{\mu\nu} = diag(1,-R^{2},-R^{2},-R^{2})$.
We assume:

(i) spatial homogeneity
$(\bigtriangledown \Phi = 0$).

(ii) $\dot{T} =\dot{S}= 0$.

\noindent
The Euler-Lagrange equations
then give
\beq
K_{i}^{k}
\ddot{\Phi}^{i}
+ 3 H K_{i}^{k}\dot{\Phi}^{i}
+ K_{ij}^{k} \dot{\Phi}^{j}\dot{\Phi}^{i}
= -V^{k}.
\label{infla}
\eeq
For inflation to occur
the first condition
to be satisfied
is that of slow rolling \ci{fluctu}
\beq
\ddot{\Phi}^{i} \ll 3H \dot{\Phi}^{i}
\label{eqB}
\eeq
and the eq. of motion becomes
\beq
3 H K_{i}^{k} \dot{\Phi}^{i} = -V^{k}
\label{eqA}
\eeq
where we have taken $3 H K_{i}^{k} \gg K_{ij}^{k}\dot{\Phi}^{j}$.
By taking the time-derivative
of eq.(\ref{eqA})
and demanding that eq.(\ref{eqB})
is satisfied, one finds the
condition \\
CONDITION  I
\beq
 \frac{V''}{V} \ll 3 K''.
\la{a1}\eeq
A second condition for
successful inflation,
besides the slow rolling
assumption, is that the energy density
is dominated by the potential.
This is equivalent to
\beq
\frac{1}{2} K_{i}^{i} \dot{\Phi^{2} }\ll V
\eeq
Using this, as well as the Einstein's
equation $3H^{2} = V$
one finds that
the condition for slow rolling becomes\\
CONDITION II
\beq
 \frac{V'}{V}  \ll \sqrt{6 K''}.
\la{a2}\eeq

\subsection{Conditions   on chiral fields due to the stability of the
dilaton
$S$}

According to the
analysis of section 3,
if the potential of the chiral matter
fields is to lead to an
inflationary
potential, it should not destabilize the dilaton solution,  spoiling
the
minimization of the dilaton and  moduli  fields.
The scalar potential is
\bea
V &=&V_0+V_1
\\
V&=&  \frac{1}{4}
e^{K}
\left(
|W-S_r W_{S}|^{2}
+ (K_{i}W+W_i)(K^{-1})^{i}_{j}(K^{j}W+W^j)-3
|W|^{2}
\right) + V_1
\nonumber\eea
with $i,j$ running over all fields but $S$ and $V_1$ the one loop
potential
\cite{axel}.
To determine the extremum of   $V$ w.r.t. $S$ we need to
solve $V_S=0$.
 In the absence of chiral matter fields the leading term in $V$ is
given by
terms proportional to $W_S$ since $W_S= - \frac{3}{2 b_0} W_0 \gg
W_0$.
Since we
do not want to spoil the stability
of $S$ due to the presence
of the chiral fields,
we have to examine
under which conditions
these terms are still dominant.  Keeping the leading terms only we
have
\bea
V_S=  \frac{1}{4}e^{K}
&[&S_r W_{SS}
(P-S_r W_{S})+ W_S[K_i(K^{-1})^{i}_{j}(K^{j}P+P^j)-3P]
\nonumber\\
&-&\frac{1}{S_r} [(K_{i}P+P_i)(K^{-1})^{i}_{j}(K^{j}P+P^j)-3|P|^2 ]
\; ]+V_{1S}.
\eea
It is now simple to see that the conditions that we need to impose on
$P(\Phi)$
and its derivatives
are:\\
CONDITION III
\beq
|S_r W_S| \gg |P|
\la{a3}\eeq
CONDITION IV
\beq
|S_r^2 W_{SS} | \gg |K_i(K^{-1})^{i}_{j}(K^{j}P+P^j)-3P|
\la{a4}\eeq
and\\
CONDITION V
\beq
|S_r^3 W_{SS} W_S| \gg
|(K_{i}P+P_i)(K^{-1})^{i}_{j}(K^{j}P+P^j)-3|P|^2|
\la{a5}\eeq
where we have taken $W=W_0(S,T)+P(\Phi,T)$.
In the case of  a single gaugino condensate $S_r W_S= -
\frac{3S_r}{2b_0}W_0$ and
$ S_r^2 W_{SS}=(\frac{3S_r}{2b_0})^2 W_0$.

A remarkable point to
observe is
that the necessity
of stabilisation of $S$ (i.e.
conditions "III, IV" and "V") imposes a constraint on the
fluctuations!

Condition "III"  gives
\beq
2 e^{-K/2}| \frac{3 S_{r}}{2b_{0}} \, m_{3/2} >  P.
\la{a6}\eeq
 One requires
$\frac{3 S_{r}}{2 b_0} \simeq 10^2$ for  reasonable solutions to the
hierarchy
problem,  i.e. a gravitino mass
$m_{3/2} =  \frac{1}{2}e^{K/2}|W|^2=1 TeV$ (or $m_{3/2}=10^{-15}$ in
natural
units)  with $e^{K/2}|=(1/S_rT_r^3)^{1/2}\simeq 1/10$ (i.e. K is
evaluated at
$\Phi=0$). Condition "III"
becomes then
\beq
2\times 10^{3}\, m_{3/2}  > P
\la{a7}\eeq
thus for a $P= \lambda\Phi^{n}$, $\Phi \simeq 1$ and $m_{3/2} = 1
TeV$
  \beq
2\times 10^{-12}  >  \lambda
\eeq
implying that the fluctuations will not be too large.

The other two constraint can be expressed in term of the gravitino
mass as
well.
Condition "IV" is
\beq
2 e^{-K/2}| \left(\frac{3S_r}{2b_0}\right)^2\, m_{3/2} \gg |
K_i(K^{-1})^{i}_{j}(K^{j}P+P^j)-3 P|
\eeq
while  "V" becomes
\beq
4 e^{-K}|  \left(\frac{3S_r}{2b_0}\right)^3 m_{3/2}^2
\gg |(K_{i}P+P_i)(K^{-1})^{i}_{j}(K^{j}P+P^j)-3|P|^2|.
\la{aa2}\eeq
Note that eq.(\ref{aa2}) sets an upper limit on the chiral potential
$V_{ch}$ with
\bea
V_{ch}
&\equiv& \frac{1}{4}
\,e^{K}\,[(K_{i}P+P_i)(K^{-1})^{i}_{j}(K^{j}P+P^j)-3|P|^2]\nonumber\\
V_{ch} &<&  e^{K} e^{-K}| \left(\frac{3S_r}{2b_0}\right)^3 m_{3/2}^2
\approx
e^{K} 10\times10^{6}\times 10^{-30}=e^{K}\,10^{-23}.
\la{aa3}\eea
We also note here that
$\lambda$ {\em
is not an arbitrary constant}.
It is  expressed as
\beq
\lambda = e^{-a T}
\la{a8}\eeq
where $a$ is a constant of order unity.
This implies that in superstring models
the small constant that is needed in order
for the inflationary fluctuations not
to be too large arises in a natural way,
and is associated with the moduli fields.  For eq.(\ref{a8}) to be
much smaller
than one it is necessary that some moduli get a large v.e.v.  This is
 the case when SUSY is broken via  a single gaugino condensate
\ci{axel}.

Let as see what
values $a$ can have:
duality invariance
implies that
\beq
\lambda =
\Pi_p \; \eta(T_p)^{-2(1+n_{ip}+
n_{jp}+n_{kp})}
\eeq
where $n_{ip}$ is the modular
weight of the pth-field w.r.t. $T_i$ and $\eta$ is the Dedekind-eta
function
($\eta(T)=e^{-\pi T/12} \Pi_n (1-e^{-2 n \pi T})$) .
Since for $T \simeq 1$, $\eta \approx 1$
we only need to consider the
$\eta$ with
large T.
If T is large $\eta
\simeq e^{-\pi T/12} $
thus
\beq
\lambda = e^{-a_i T_i}
\eeq
with
\beq
a_i = -\frac{\pi}
{6}(1+n_{ip}+n_{jp}+n_{kp}).
\eeq
If we consider a $\Phi^3$ term so that all
three fields have the same modular
weight
\beq
a_i = - \frac{\pi}{6} (1+3n_i).
\eeq
For an overall moduli ($T_i=T_j$) and
$N=3n_i$,
$\Sigma_{i=1}^3 a_i=\pi  (1+N)/2$.
For untwisted fields
$N=-1$ and $a=0$. We
note that $a$ could vary if the Yukawa
coupling $\lambda$ is
multiplied by a modular invariant
function becoming thus  model
dependent.

\subsection{Region with small $|\Phi|$}

We initially consider the
limit with $|\Phi| \ll 1$.
In this case
the leading term is either linear
or quadratic in $\Phi$.
In the absence of a
linear term in $W$,  $V$  has no linear term
and the quadratic one will be  the dominant.
In \cite{axel} was shown that
all scalar fields acquire a mass
$m_{0} > 0$ and therefore
at the Plank scale $\Phi=0$ is a stable solution\footnote{ At lower
scales
$\Phi$ may acquire a v.e.v. $\neq 0$.}.
Then
\beq
V \approx m_{0}^{2}|\Phi|^{2}
\eeq
\beq
\frac{\partial V}{\partial \Phi} = m_{0}^{2}\Phi.
\eeq
Thus
\beq
N = -\int \frac{K_{\Phi}^{\Phi}V}{V_{\Phi}}d\Phi
= -\int K_{\Phi}^{\Phi}\Phi d\Phi
\eeq
For $\Phi \ll 1$, $K_{\Phi}^{\Phi} = \frac{1}{T^n_{r}}$
thus
\beq
N = -K_{\Phi}^{\Phi}|\Phi|^{2} = \frac{|\Phi_{i}|^{2}}{T^n_{r}} \ll
1.
\eeq
This indicates that in the regime
$\Phi \ll 1$ there is no inflation.
This effect comes from the smallness
of $K_{\Phi}^{\Phi}$,
which suppresses the number of e-folds.

However, $\Phi$ can be much larger.
In fact, for the usual kinetic term for $\Phi$
\beq
K = -\ln(T^n_{r}-|\Phi|^{2}), \; \; \;  \; \; n = 1, 2, 3
\eeq
the condition is that
\beq
T^n_{r} - |\Phi|^{2} < 1.
\eeq
Then if $T^n_{r}$ is large we can have
$|\Phi|^{2} \sim T^n_{r} >1$.
In \cite{axel} it was found that
$ T_{r} \simeq 17, 24, 44$
depending on how many moduli acquire a large v.e.v.

In this case
\begin{eqnarray}
\frac{\partial K}{\partial \Phi} & = &
\frac{\Phi}{T^n_{r}-|\Phi|^{2}},
 \nonumber \\
\frac{\partial K}{\partial \Phi \; \partial \bar{\Phi}} & = &
\frac{|\Phi|^{2}}{(T^n_{r}-|\Phi|^{2})^{2}}.
\end{eqnarray}
Both these quantities can be larger than one  for
$T^n_{r} \simeq |\Phi|^{2} > 1$,
therefore
the number of e-folds will not be
necessarily suppressed, as in the previous case.

In the following
sections, we will examine
the possibility of inflation in
this regime. We stabilize the dilaton
with a single gaugino condensate, but the results
also apply
to the models which minimize the dilaton by using two gaugino
condensates. An advantage of \cite{axel}
 is that by fixing $S$, $T$ is also stabilized
and that $T$ may acquire large vev's, allowing for a larger
$\Phi$. In the case of
two gaugino condensates the arguments
work in a similar way,
once $T$ is also minimized ($T \simeq 1.2$).

\subsection{ Large $|\Phi|$
region - generic behaviour}

The tree level scalar potential  (cf. eq(\ref{eq1})) is
\beq
V =  \frac{1}{4} e^{K} f
\eeq
withu100
\begin{eqnarray}
f& = &
G_{i}(K^{-1})^{i}_{j}G^{j} - 3|W|^{2} \nonumber \\
 & = &
(K_{i}W+W_{i})(K^{-1})_{j}^{i}
(K^{j}W+W^{j})-3|W|^{2}.
\la{aa1}\end{eqnarray}
The derivative with respect to $\phi_i$
which, we  denot by $'$,  is
\beq
V' =  \frac{1}{4} e^{K}(K'f+f').
\eeq
There are three possibilities:

(i) $K'f \gg f'$

(ii) $K'f \ll f'$

(iii) $K'f \simeq f'$\\
In the former case
\beq
V' \simeq  \frac{1}{4} e^{K}K'f
\eeq
thus
\beq
N = -\int \frac{K''V}{V'}d\Phi
= -\int \frac {K''}{K'} d\Phi
\la{a9}\eeq
and for $ K = -\ln(T^n_{r}-|\Phi|^{2})$
\beq
N = \ln \left (
\frac{|\Phi|^{2}}{T^n_{r}-|\Phi|^{2}}\right).
\la{a10}\eeq
To get inflation from this solution, one
would require too much fine tuning,
therefore the scheme is not viable.

In the second case
\beq
V' \simeq  \frac{1}{4}e^{K}f'
\eeq
thus
\beq
N = -\int \frac{K''V}{V'}d\Phi =
-\int \frac{K''f}{f'}d\Phi.
\eeq
During inflation $f>0$
and we will consider the space where $f$ does not change sign.
Then
\beq
|N| = \left |-\int \frac{K''f}{f'}d\Phi\right|
< \left|-\int \frac{K''}{K'}d\Phi\right|
\eeq
and we know from the previous case that
this function does not
lead to enough e-folds.
This indicates that,
for a potential to inflate, there has
to be some cancellation between
$K'f$ and $f'$.

Let us also remember that
a usual problem in dynamical symmetry
breaking is the existence of a
negative
cosmological constant. In supergravity models where SUSY is broken
dynamically via
gaugino condensation the scalar potential is typically $V\simeq
-O(\Lambda_c^4)$. It is not clear how the  cosmological constant will
vanish. A
possible solution is to introduce a linear superpotential $P = cD$
where
$D$ is a
chiral field and $c$ is a constant to be fine tuned to give zero
cosmological
constant \cite{cos}.  With the inclusion of $cD$ in the
superpotential the
dominant term comes from $|P_D|^2=|c|^2$ which is indeed positive and
allows
for the cancellation of the vacuum energy.  Minimizing with respect
to
$D$ one
obtains $cD\simeq W_0 \neq 0$. In the absence of a better way to
cancel the
cosmological constant we will assume the existence of this term,
however, the
main point is not how the cosmological constant is canceled but that
for
vanishing v.e.v. of the chiral fields $\Phi$ the minimum of $V$ is at
$V=0$ \cite{cos}.
Therefore, the  inflationary potential is given by those
terms
which are different than zero for $\Phi \neq 0$ and do not
destabilize $S$.

\section{Viable solution-general description}

In this section we will give a viable solution that leads to an
inflationary
potential and has the right magnitude
of density fluctuations. We will consider the chiral scalar
potential induced
by the chiral superpotential $P(\Phi,T)$ and do not spoil the minimum
of the
dilaton field.

Let us study the region where $\Phi >1$ and
\beq
K'=\frac{\partial K}{\partial \Phi} =
\frac{\Phi}{T^n_{r}-|\Phi|^{2}} >1.
\eeq
In this region one expect  the leading term of
the scalar chiral potential
\beq
V_{ch} =  \frac{1}{4} e^{K}
\left [
(K^{-1})_{b}^{a}
(K_{a}P+P_{a})(K^{b}P+P^{b})-3|P|^{2}
\right ]
\eeq
to be the one proportional to $|P|^2$. So, let us take  for
simplicity and
illustration  purposes the potential
\beq
V_{ch}=e^{K}\,A\,|P|^2.
\eeq
Taking the derivative with respect to $\Phi$
\bea
V'_{ch}&=&e^K A\, (K' |P|^2+\bar{P}P')
\nonumber\\
V'_{ch}&\simeq&e^K A\, K' |P|^2
\eea
where we have  kept the leading term only.
It is easy to see that the number of e-folds of inflation will not be
enough
since
$K''\,V/V'=K''/K'$ (cf. eq.(\ref{a10})).  Therefore, an inflationary
potential
requires $V$ not to be dominated by $|P|^2$. One of the simplest
possible
choices is to introduce two chiral fields and  consider the trilinear
superpotential
\beq
P = \lambda(T) (\Phi_{1}^{3}-\Phi_{2}^{3}).
\la{a11}\eeq
The derivative with respect to $\Phi_2$ gives
\beq
\frac{\partial V_{ch}}{\partial \Phi_{2}} =  \frac{1}{4}
e^{K} ({K_{2}f+f_{2}})
\eeq
which to leading order is
\beq
V_{ch, 2} \sim  \frac{1}{4}
K_{2}|P|^{2}(K_{a}(K^{-1})^{a}_{b}K^{b}-3)=0.
\la{a12}\eeq
$V_{ch, 2}$ vanishes\footnote{
In a subsequent section we calculate the
derivatives of $K$ with respect to the fields
and it will become obvious that
$K_{a}(K^{-1})^{a}_{b}K^{b} \neq 3$.}  if $P \simeq 0$, i.e. $
<\Phi_{1}>
\simeq <\Phi_{2}>$. Furthermore, the most stringent condition on $P$
or
$P_i$ comes
from condition III eq.(\ref{a3}) on the superpotential $P$.

We will work then in the region $P\simeq 0$ ($  <\Phi_{1}>
\simeq <\Phi_{2}>$) and
$P_i \gg W_0, cD$ while still satisfying the S-stability conditions
"III, IV,
V". We see that there is indeed an allowed region where $W_S >
P_1=3 \lambda \Phi_1^2
> W_0, cD$.

\subsection{Scalar potential
and derivatives}

Let us consider the two chiral  fields introduced in sec.5  belonging
to oneu40
sector of the orbifold only. In this case the modular weights of
these fields
will be different than zero only with respect to one of the moduli
fields
$T_i$, say $T_1$, and with a  K\"{a}hler potential
\beq
K_0 = -\ln Q-\ln{T_{r2}}-\ln{T_{r3}}, \; \; \;\;\; Q \equiv
T_{r_{1}}-|\Phi_{1}|^{2}
- |\Phi_{2}|^{2}
\la{a13}\eeq
and
\begin{eqnarray}
K_{i} & = & \frac{\partial K}{\partial \Phi_{i}}
= \frac{\bar{\Phi}_{i}}{Q} \nonumber \\
K_{T_1} & = & - \frac{1}{Q}
\end{eqnarray}
where the index $i=1,2$ denotes the two $\Phi$
fields which we are using.
Similarly one easily calculates the
second derivatives of $K$ with respect to all field
combinations, to form $K$ (in obvious notation)
\begin{equation}
(K^a_b)=
\left(
\begin{array}{ccc}
K_{T}^{T} & K_{T}^{1} & K_{T}^{2}
\\
K_{1}^{T} & K_{1}^{1} & K_{1}^{2}
\\
K_{2}^{T} & K_{2}^{1} & K_{2}^{2} \\
\end{array}
\right) =
\frac{1}{Q^{2}}
\left(
\begin{array}{ccc}
1 & -\Phi_{1} & -\Phi_{2}
\\
-\bar{\Phi}_{1} & T_{r}-|\Phi_{2}|^{2}
& \bar\Phi_{1}\Phi^2
\\
-\bar{\Phi}_{2} & \bar\Phi_{2}\Phi^{1} & T_{r}-|\Phi_{1}|^{2}
\\
\end{array}
\right).
\la{a14}\end{equation}
The inverse matrix is then
\begin{equation}
(K^{-1})_{i}^{j} =
Q \left(
\begin{array}{ccc}
T_{r} & \Phi_{1} & \Phi_{2}
\\
\bar{\Phi}_{1} & 1 & 0
\\
\bar{\Phi}_{2} & 0 & 1
\\
\end{array}
\right).
\la{a15}\end{equation}
The moduli $T_2, T_3$ do not have any mixing terms with $T_1$ and
$\Phi_i$ in
$(K)$ or its inverse and we have not included them in
eqs.(\ref{a14}) and
(\ref{a15}).  It is easy to see that $K_{a}(K^{-1})^{a}_{b}K^{b}=4$
where $a,
b=S,T_i,\Phi_i$.

To determine the number of e-folds of inflation and the density
fluctuations we
need to calculate the potential and
its first and second derivatives.  As shown in eq.(\ref{a12}) to
leading order
the solution to $V_2=0$ is
\beq
< \Phi_{1}>  =  < \Phi_{2}>
\eeq
which implies
\bea
<P> &=& 0\nonumber\\
<P_{T}> & = &  \frac{\lambda_{T}}{\lambda} <P> =0
\eea
and
\beq
<P_{1}> =  -< P_{2}>.
\nonumber\eeq
Using eqs.(\ref{aa1}), (\ref{a11})  and (\ref{a13}) we find that
\beq
<f> = 2 (K^{-1})_{1}^{1}| P_{1}|^2.
\eeq
Similarly
\begin{eqnarray}
<f_{1}> & = &
(K^{-1})_{1}^{1}\bar{P}^{1}
P_{11} + 2 (K^{-1})^{1}_{11}
|P_{1}|^{2}
\nonumber \\
<f_{11}> & = &
(K^{-1})_{1}^{1}\bar{P}_{1}
P_{111} + 2 (K^{-1})^{1}_{11}
P_{11}\bar{P}^{1}
+ 2 (K^{-1})^{1}_{111}
|P_{1}|^{2}.
\end{eqnarray}
 Using these eqs. it is easy to show that
the scalar potential
is simply given by
\beq
V =  \frac{1}{2} e^{K}
(K^{-1})_{1}^{1} |P_{1}|^2
\la{a16}\eeq
with
\beq
P_1=3\lambda\Phi_1^2,\;\;\;\;\;(K^{-1})_{1}^{1}=Q
\eeq
while
\bea
V_{1} &=& \frac{1}{4}e^{K}( K_{1}f + f_{1})
\nonumber\\
<V_{1}> & = &  \frac{1}{4}e^{K}
(2|P_{1}|^2[ K_1 (K^{-1})_{1}^{1} + (K^{-1})_{11}^{1} ] +
(K^{-1})_{1}^{1}\bar{P}^{1}P_{11})
\la{a17}\\
<V_{1}> & = & \frac{1}{4}e^{K}
(K^{-1})_{1}^{1}\bar{P}^{1}P_{11}
\nonumber \eea
and
\bea
V_{11}  &=& \frac{1}{4}e^{K} (K_{11}f
+2K_{1}f_{1}+K_{1}^{2}f+f_{11})\\
<V_{11}> & = &  \frac{1}{4}e^{K}(K^{-1})_{1}^{1}\bar{P}^{1}P_{111}.
\la{a18}\nonumber\eea
Therefore,
\beq
\frac{<V>}{<V_{1}>} = \frac{2P_{1}}{P_{11}}= \Phi_{1}
\la{aa18}\eeq
\beq
\frac{<V>}{<V_{11}>} = \frac{2P_{1}}{P_{111}}= \Phi_{1}^{2}
\eeq
and the conditions for
successful inflation "I,II"  given in eqs.(\ref{a1}) and (\ref{a2})
are satisfied. In fact, for our potential if eq.(\ref{a1}) is
satisfied so will
be eq.(\ref{a2}) and the value of $\Phi_i$ that breaks the condition
is
$|\Phi_{1 \,end}|^2=T_{r_{1}}(\frac{1}{2}-\frac{\sqrt{21}}{14})$.

Note that in eq.(\ref{a17}) the term proportional to $|P_1|^2$
vanishes. This
fact is important  for having  enough inflation.  In the limit where
$P\simeq 0$
a necessary  condition on $K$
is that the term proportional to $P_i^2$ in $V_1$ is suppressed with
respect to
the one in $V$, i.e.
\beq
P_iP^j (K^{-1})_{j}^{i} \gg P_iP^j (K_1 (K^{-1})_{j}^{i}
+(K^{-1})_{j1}^{i}),
\eeq
so that  their is some cancellation between $K'f $ and $f'$.
This is exactly what happens in our example where $(K_1
(K^{-1})_{j}^{i}
+(K^{-1})_{j1}^{i})=0$.

\subsection{Number of e-folds and fluctuations}

We now calculate  the
predictions for the number of
e-folds and the fluctuations for our example.

For two fields we have the following
starting equation:
\beq
3 H K_{1}^{1} \dot{\Phi}_{1}
+ 3 H K_{2}^{1} \dot{\Phi}_{2}
= V_{1}.
\eeq
However
$\Phi_{1}$ and $\Phi_{2}$ have
identical contributions to
the scalar potential and the
kinetic energy, as well as
the same initial conditions.
For this reason,
to quantify our
results we can make the
assumption
$\dot{\Phi}_{1} = \dot{\Phi}_{2}$
and
$\Phi_{1} = \Phi_{2}$.
In addition we take the fields to be
real and then we
find that the number of e-folds
is given by the equation
\beq
N =  -\int
\left
[\frac{V}{V'}
(K_{1}^{1}+K_{2}^{1})
\right
] d\Phi_{1}
\eeq

\beq
N =  -\int
\left [
(K_{1}^{1}+K_{2}^{1})
\Phi_{1}
\right ] d\Phi_{1}
\eeq
where
\beq
(K_{1}^{1}+K_{2}^{1})
= \frac{T_{r_{1}}}{Q^2}=\frac{T_{r_{1}}}
{(T_{r_{1}}-2|\Phi|_{1}^{2})^{2}}.
\la{a19}\eeq
Then
\beq
N =\frac{T_{r_{1}}}{4Q}=
\frac{T_{r_{1}}}{4(T_{r_{1}}-2| \Phi_{1,init}|^{2})}
\la{a20}\eeq
where the subindex ``{\em init}''
refers to the initial value of the
field.

If inflation
occurs at a scale
$10^{17}$ GeV, for enough growth of the scale
factor $N$ has to be
$ \geq 65$. On the other hand if inflation occurs at a lower
scale, the number of e-folds required is smaller:
for example, the scale at which an $S$-dependent superpotential
arises, through gaugino condensation, is of the order
of $10^{12}$ GeV. At this stage the region of the universe
which today is within the horizon was composed by
$10^{23}$ causally disconnected domains. For enough inflation,
we would now need $N$ to be $\geq 53$
and we see that we achieve this
for a reasonable choice of
$T_{r_{1}}$ and $\Phi$.
For $\Phi \simeq \sqrt{T_{r_{1}}/2}$,  $Q$
will be
small
allowing for  $N \ge 53$.  Note that the value of $T_r$ can be
$(17,24,44)$
depending on how many moduli get a large v.e.v. and these values have
been obtained
dynamically.
Another thing to note is that the value
of $\Phi_{1,2}$ at the end of inflation
is not so relevant for the number of e-folds,
as what matters is just the initial conditions.
However it is easy to see
from the conditions "I"  and "II"
that inflation will indeed
come to an end,
as $K''$ decreases
with decreasing $\Phi_{1,2}$,
and at some stage
the conditions will break down (for $|\Phi_{1
\,end}|^2=T_{r_{1}}(\frac{1}{2}-\frac{\sqrt{21}}{14})$). So the
allowed
region for
inflation is $T_{r_{1}}/2 > |\Phi_1|^2>
T_{r_{1}}(\frac{1}{2}-\frac{\sqrt{21}}{14})$.

We now procede to calculate the
fluctuations $\delta \rho / \rho$ that
appear in the model.
In general, the computation of density perturbations
with more than one inflaton field
is a highly non-trivial problem, in the general case
\cite{twoi}.
However, in our model, inflation occurs for
two scalar fields which appear in exactly the same
way in the equations of motion and also have similar
initial values. In this case, the
equations of motion are significantly simplified,
since the symmetries of the model allow us to
consider the two fields in an equivalent footing.
The other fields of the theory (dilaton and moduli) are
already frozen at this stage so no additional complications
occur.
Then, instead of having to solve a system of
coupled equations, we can read the magnitude of
the fluctuations by simply looking at one of the
two identical equations of motion.
Taking into account that the dilaton and the
moduli fields are frozen, we can read
$E_{kin,tot}$  from the equation for
the Largangian density of the system
(\ref{lag}) to be
\beq
E_{kin,tot} = \frac{1}{2} K_{1}^{1}
\dot{\Phi}_{1}^{2}
+ \frac{1}{2} K_{2}^{2}\dot{\Phi}_{2}
+ \frac{1}{2}(K_{1}^{2}+K^{1}_{2}) \dot{\Phi}_{1}
\dot{\Phi}_{2}
\la{ekin}\eeq
where the only modifications from the
usual terms that one would expect,
appears due to the fact that in superstring
models the kinetic terms are
non-canonical.

The fluctuations of the scalar fields
$\Phi_{1}$ and $\Phi_{2}$
that are produced during inflation are equal and
lead to the density inhomogeinities \ci{fluc}
\beq
\frac{\delta \rho}{\rho}
=
C \, \left. \frac{H(\Phi) \delta \Phi}{ \dot{\Phi}}
\right|_{k \sim H}
\eeq
where the constant $C$ is of order 1 and we took $\Phi=\Phi_1=\Phi_2$. For a
hot or cold universe  one has  $C=-4/3$ and $C=-6/5$ respectively
\ci{C.hotcold} and $\delta \Phi  = \frac{H}{2 \pi} [(K_{1}^{1} +
K_{1}^{2})]^{-1/2}$ since the kinetic term is non canonical (cf.
eq.(\ref{ekin})).
Using
\beq
\dot{\Phi}_{1} =
\frac{V_{1}}{3 H (K_{1}^{1}+K_{1}^{2})}
\eeq
we find that the fluctuations
scale as
\beq
\frac{\delta \rho}{\rho} =  \frac{C}{ 2 \;\pi
\sqrt{3}}
\sqrt{K_{1}^{1}+K_{1}^{2}}
\frac{V^{3/2}}{V_1} .
\la{a21}\eeq
and defining
$E(\Phi) \equiv V/V'$ eq.(\ref{a21}) becomes
\beq
\frac{\delta \rho}{\rho} =  \frac{C}{2 \pi
\sqrt{3}}
\sqrt{K_{1}^{1}+K_{1}^{2}}\, E\,
V^{1/2}
\la{a22}\eeq
We can find an upper limit to the fluctuations using the
constraint on
$V_{ch}$ given in eq.(\ref{aa3})
\beq
\frac{\delta \rho}{\rho} <  \frac{C}{2 \pi
\sqrt{3}}
\sqrt{K_{1}^{1}+K_{1}^{2}} \,E
\,e^{K/2} e^{-K/2}| \left(\frac{3S_r}{2b_0}\right)^{3/2} m_{3/2}
\la{b1}\eeq
where $e^{K}|$ is evaluated at $\Phi=0$. For a large hierarchy
solution
$\left(\frac{3S_r}{2b_0}\right)^{3/2} m_{3/2} \approx 10^{-12}$ and
eq.(\ref{b1}) gives
\beq
\frac{\delta \rho}{\rho} <  \frac{C\;10^{-12}}{2 \pi
\sqrt{3}}
\sqrt{K_{1}^{1}+K_{1}^{2}}
\,e^{K/2} e^{-K/2}| \;E
\la{b2}\eeq
The density fluctuations will, in general, be much smaller
than the ones observed by COBE \ci{COBE}  where $\frac{\delta \rho}{\rho}
\simeq 10^{-5}$. This is a known feature for a chaotic potential with $V^{1/4}
\sim 10^{13} GeV$ and $E=V/V' \sim \phi =O(1)$ which is precisely our case.
 Larger density fluctuations can be obtained  if the potential is very flat.
The fluctuations  will be of the right order of magnitude   if
$E=V/V'=O(10^{3-7})$ as can be seen from eq.(\ref{b2}), since we expect
$\sqrt{K_{1}^{1}+K_{1}^{2}}e^{K/2} e^{-K/2}| \sim O(10^{1-5})$. Such a
potential may appear naturally in hybrid models \ci{hybrid}.
Furthermore, other sources of fluctuations, like isothermal fluctuations
\ci{isothermal} or vacuum fluctuations of the electromagnetic field
\ci{Veneziano},  can give significant contributions $\delta \rho /\rho$.
Finally, several
additional sources of fluctuations, like those due to
the existence of topological defects
are to be expected in the early universe.

For the potential described in eq.(\ref{a16}), (\ref{a17}) we have
\bea
\sqrt{K_{1}^{1}+K_{1}^{2}}e^{K/2} e^{-K/2}| &=&
\frac{\sqrt{T_{r1}}}{Q}\;\frac{1}{\sqrt{T_{r2}T_{r3}Q}}
\;\sqrt{T_{r1}T_{r2}T_{r3}}\nonumber\\
&=& \frac{T_{r1}}{Q^{3/2}}=\frac{(4N)^{3/2}}{\sqrt{T_{r1}}}
\la{}\eea
 where we used eq.(\ref{a20}). The value of $N$ is determined
by the horizon scale today. In fact,
for a fluctuation emitted with a certain wavelength
during inflation, one may
calculate the wavelength
that the fluctuation has
today and compare this to the horizon
distance $(6000 \; {\rm Mpc})$
\cite{fluctu}. Indeed, during inflation
a wave emitted at some
value $\Phi_{1}$
increases its wavelength.
Taking into account that
the universe reheats at a temperature
$T_{reheat}$
and subsequently cools to a temperature
$T_{\gamma} \sim 3 K$,
we find that scales of the order
of the horizon today correspond
to fluctuations emitted around $65$
e-folds before the end of inflation.
The exact time of emission
of a fluctuation that corresponds to the
present horizon depends on the value of
$T_{reheat}$, which is model dependent.
We look at this
issue, in connection with the gravitino problem
as well, in future work.

 For $T_{reheat}=V^{1/4}_{end}$
 we have $N=65$ which gives  $\sqrt{K_{1}^{1}+K_{1}^{2}}e^{K/2} e^{-K/2}| \sim
10^{3}$ and using (\ref{a13}), (\ref{a19})  the fluctuations (cf.
eq.(\ref{a22})) are
\beq
\frac{\delta \rho}{\rho} =  \frac{C}{2 \pi
\sqrt{3}}
\frac{\sqrt{T_{r_{1}}}}{Q}
\Phi
V^{1/2}
\eeq
with
\beq
V^{1/2}=\frac{1}{2} e^{K/2} \sqrt{2 (K^{-1})_{1}^{1}}\;\;P_1.
\eeq
The  upper limit on the fluctuations in our example is
\bea
\frac{\delta \rho}{\rho} &<& \frac{C }{2 \pi \sqrt{6}}
\left(\frac{T_{r_{1}}}{Q}\right)^{3/2}
\left(\frac{3S_r}{2b_0}\right)^{3/2} m_{3/2}
\nonumber\\
\frac{\delta \rho}{\rho}&<&\frac{C\, 10^{-12}}{2\pi \sqrt{6}}
(4N)^{3/2} \sim 10^{-10}
\la{a23}\eea
In this last eq. we  used eq.(\ref{a20}), $m_{3/2}= 1\,TeV,
\left(\frac{3S_r}{2b_0}\right)=10^2$,
$<|\Phi_{i}|^2>\simeq
T_{r_{1}}/2, i=1,2$ and $N=65$. The fluctuations in eq.(\ref{a23}) are clearly
too small to explain the temperature inhomogeneities observed by COBE. They
could be slightly increased if   we generalize the potential $V$ to have
$e^{-K}=Q^3$
(i.e.
$Q_2=Q_3=Q_1=Q$ cf. eq.(\ref{a13})) and the same superpotential of
eq.(\ref{a11}). However, the fluctuations are still too small since
$\sqrt{K_{1}^{1}+K_{1}^{2}}e^{K/2} e^{-K/2}|
=\frac{T_{r}^2}{Q^{5/2}}=\frac{(4N)^{5/2}}{\sqrt{T_{r}}}\sim 10^{5}$ and
$\frac{\delta \rho}{\rho}\sim 10^{-8}$. As mentioned above,  the fluctuations
will be larger if the potential is very flat as in hybrid models. They are also
amplified when taking into account the
contibutions of isothermal fluctuations.
Finally, the bounds we obtain are valid in this particular scheme, where
we stabilize the dilaton by the formation
of a gaugino condensate at a rather low scale. In alternative schemes,
which we discuss in future work, an S-dual superpotential
results to the stabilisation of the dilaton and therefore to the
possibility for inflation at a much earlier stage. In this class of
solutions the density fluctuations tend to be larger.

We would like to emphasize the following:

$\bullet$
The values we used  for  $S$ and $T$
are {\it not arbitrary}; they are
dynamically  obtained by minimizing the scalar potential in the
absence of
chiral matter fields.
As an example
\ci{axel} we
can take an $SU(6)$ gauge group in the hidden sector with 6 chiral
fermions
(i.e. $b_0=15/16\pi^2$). For this example the value of the dilaton is
$S_r=2Re\,S=4.2$ and the moduli fields get a v.e.v. of
$T_{ri}=(17,24,44)$
depending on
whether there are 3,2 or 1 moduli with large v.e.v. (the
other moduli
get a v.e.v. $|T|=1$).
It is for these realistic field values, that the density fluctuations
are constraint to be below the
limits we have mentioned.

In all cases, we expect
the fluctuations to be close to their upper
limit, not only because the larger overdensities dominate
over any source of smaller distortions, but mainly
because  the potential that leads to the
largest fluctuations is exactly the one for which
the conditions for successful inflation are first
satisfied. Since we have only
one inflationary era, the spectrum of fluctuations
we predict is the usual scale-invariant
Harrison-Zeldovich one. This differs from
other models which use two scalar
fields, one light and one heavy, in order to obtain
two stages of inflation.
Obtaining a non-trivial spectrum of fluctuations
is also possible in our approach, by the introduction of
more fields. This is addressed in future work.
Here, however, we wanted
to look at the simplest possible models that are
consistent with inflation. Our basic aim was to point out
that it is possible for chiral fields to act as inflatons,
provided that the dilaton and moduli fields are
frozen and that the requirement that the potential of the
chiral fields does not destabilise the dilaton
minimization leads to interesting constrains
for the density fluctuations, in terms of quantities
which are dynamically determined.

\section{Summary and Conclusions}

        We studied the possibility of having an inflationary
potential in 4-D string
models obtained from the heterotic string compactified on orbifolds.
A generic
feature of all these string vacua is  the existence of the dilaton
$S$ and
moduli $T_i$ fields. The interaction between $S$ and the other scalar
fields
is generic and has a standard form in the context of gaugino
condensation. We
found  that it is not possible
for  the
potential to inflate in  the $S$ direction and since the potential
must inflate
for  all dynamical fields we conclude that the dilaton should be at
its
minimum.  We studied  the stabilisation of the dilaton field  in the
context of
 gaugino condensation  which generates an effective $S$-dependent
superpotential. The moduli fields share some of the problems of the
dilaton,
however their dynamics is more cumbersome. Nevertheless, in the
context of a
single gaugino condensate, the moduli and the dilaton fields are
frozen at the
same time and their v.e.vs break supersymmetry. We would then expect
an
inflationary potential only below scale where
a gaugino condensation may form.
If a gaugino condensation forms in a $N=1$
supergravity theory, supersymmetry will be broken.
For reasonable values of the masses of the supersymmetric scalar
fields, the scale where gaugino condensation occurs is  around
$10^{12}\,GeV$).
The picture that we have is then a universe that starts with random
values of
the different fields (dilaton, moduli, chiral matter fields). The
universe
cools down and it  evolves in a standard non-inflationary way
until the $S$
and $T$ are stabilized. Below this scale, other fields, like the
chiral matter
fields,  could drive an exponentially fast expansion of the universe
as long as
its potential does not destabilize $S$ and $T$. So,  we expect that
the
universe  arrives at an  inflationary period naturally when the
fields roll
down to their minimum and the inflationary conditions are met.  We
have studied
under which conditions this is possible
and we have shown a simple example.
Furthermore, we have found that
the density fluctuations due to the inflaton field are
constraint
by the above
dynamics.

To be more precise, we have found that

(i) \, the existence and way of coupling
of the dilaton $S$ does not allow for inflation,
unless $S$ is stabilized.
Before this occurs,
no other field may be used as
the inflaton.

(ii) \, At the same time, the
moduli fields are also stabilised, and may
acquire large vev's.

(iii) \,
Any potential with
$V|_{X_{r_{0}}} = V'|_{X_{r_{0}}} = 0$
and $ K_{X}^{X} = \frac{1}{X^{2}}$
(like in the case with $X\equiv S$)
will not inflate around the minimum.

(iv) \,
For the case
$ K_{X}^{X} = \frac{1}{X^{2}}$
and $V \simeq e^{-b X}$, or
$V = \frac{1}{X}$, no inflation occurs either.

(v) \,
However, the dynamical evolution of
chiral fields leads to inflation.
This occurs
in the region with $|\Phi_{i}|^{2} \simeq T^n_{r}$,
while if $|\Phi_{i}|^{2} \ll T^n_{r}$,
it is not possible to obtain enough
e-folds of inflation.

(vi) \,
For a potential
$V = e^{K}f$, $V' = e^{K}
(K'f+f')$, the regions
$K'f \gg f'$ or $K'f \ll f'$
may not inflate. For sufficient
growth of the scale factor,
there has to be a cancellation
between $K'f$ and $f'$.

(vii) \,
It is remarkable
that, in order for the potential of $\Phi_{i}$
not to destabilize the dilaton
minimization,
a constraint that leads to an
{\em upper limit on the
magnitude of the
fluctuations}
arises.

(viii) \,
In this framework,
a simple superpotential of the
form $W = \lambda(\Phi_{1}^{3}-
\Phi_{2}^{3})$, may inflate for
values of $T$
that have been derived independently
of this calculation.
The resulting fluctuations in the simplest
scheme are smaller than those measured by COBE.

\vspace{0.5 cm}

\noindent
{\bf Acknowledgment }

\noindent
We would like to thank  G. G. Ross for many enlightening discussions
and comments and
for collaborating  at  the early stages of
this work.

\end{document}